\documentclass[10pt,a4paper]{article}
\usepackage[ascii]{inputenc}
\usepackage[]{fontenc}
\usepackage[english]{babel}
\usepackage{amsmath,amssymb,amsfonts,textcomp}
\usepackage{color}
\usepackage{array}
\usepackage{hhline}
\usepackage{hyperref}
\hypersetup{pdftex, colorlinks=true, linkcolor=blue, citecolor=blue, filecolor=blue, pagecolor=blue, urlcolor=blue, pdftitle=, pdfauthor=, pdfsubject=, pdfkeywords=}
\usepackage[]{graphicx}
\makeatletter
\newcommand\ps@Standard{
  \renewcommand\@oddhead{}
  \renewcommand\@evenhead{}
  \renewcommand\@oddfoot{}
  \renewcommand\@evenfoot{}
  \renewcommand\thepage{\arabic{page}}
}

\makeatother
\title{Galaxies and cladistics}

\author{Didier Fraix-Burnet}
\begin{document}

\maketitle

Laboratoire d{\textquotesingle}Astrophysique de Grenoble, UMR5571
CNRS/Universit\'e Joseph Fourier

didier.fraix-burnet@obs.ujf-grenoble.fr

\bigskip

\begin{center}
 Original text published in \\
 \href{http://www.springer.com/life+sci/book/978-3-642-00951-8}{\textit{Evolutionary Biology. Concept, Modelization and Application}}, 2009, Pontarotti, P. (Ed.), Springer, pp. $363-378$ 
 DOI~10.1007/978-3-642-00952-5
\end{center}

\bigskip

{\bfseries
Abstract}

The Hubble tuning fork diagram, based on \index{morphology}morphology
and established in the 1930s, has always been the preferred scheme for
classification of galaxies. However, the current large amount of
multiwavelength data, most often spectra, for objects up to very high
distances, asks for more sophisticated statistical approaches.
Interpreting formation and evolution of galaxies as a
{\textquotesingle}transmission with modification{\textquotesingle}
process, we have shown that the concepts and tools of phylogenetic
systematics can be heuristically transposed to the case of galaxies.
This approach that we call
{\textasciigrave}{\textasciigrave}astrocladistics{\textquotesingle}{\textquotesingle},
has successfully been applied on several samples. Many difficulties
still remain, some of them being specific to the nature of both
galaxies and their diversification processes, some others being
classical in cladistics, like the pertinence of the descriptors in
conveying any useful evolutionary information.

\section{1. Introduction}
\index{Galaxies}Galaxies have been discovered quite recently with
respect to the History of Humanity, since it was in 1922 by Edwin
Hubble. Our own galaxy, called the Milky Way, was rather well known at
this epoch. Hubble found that there are indeed many other similar
objects situated at very far distances from us, much farther away than
the dimension of the Milky Way. Galaxies are now known to be
fundamental entities of the Universe and can be defined as
self-gravitating ensemble of stars, gas and/or
{\guillemotleft}~dust~{\guillemotright} (dust being mainly grains for
astronomers). 

Typical big galaxies like our own are about 100 000 light-years across.
Shapes are basically those defined by Hubble, that is spiral,
elliptical and irregular. This reflects the orbits of the stars that
can be distributed in a disk (spirals), in a regular three dimension
structure (ellipticals), or in a more disturbed fashion (irregulars).
In spiral galaxies, density waves create spiral arms, sometimes with a
bar-like structure around the centre of the galaxy. These density waves
are not necessarily long-lasting, but they tend to concentrate gas and
dust so that star formation is mostly associated with them. 

Self-gravitating objects have probably appeared early in the Universe
(which age is now established to be 13.7 10\textsuperscript{9} years),
in association with the very slightly inhomogeneous distribution of
gravitational matter. The structure of the Universe is shaped by the
dark energy (70\% of the Universe total energy), that causes the
expansion, and the gravitational field that is fashioned essentially by
the dark matter (26\%). Galaxies are made up of baryonic matter which
represents only 4\% of the composition of the Universe. The primeval
inhomogeneities of the gravitational potential are observed at the
recombination epoch (time when the Universe became transparent at an
age of about 400 000 years) thanks to their thermal radiation that the
expansion of the Universe has now
{\textquotedblleft}cooled{\textquotedblright} down to 3K (e.g. Hinshaw
et al. 2009)). These early overdensities have grown up by condensation
to form bigger and bigger structures called halos and filaments. The
baryonic matter is gravitationally affected by the evolution of the
dark matter distribution, and collapses of small scale entities created
the first light emitting objects (first stars at about 400 millions
years) whose distribution can be observed now in the infrared
(Kashlinsky et al. 2005). Currently, most distant galaxies can be seen
at redshifts about 6 or 7, equivalent to distances of about 12 Gyr or
90\% of the age of the Universe in lookback time. Since the Universe is
in expansion, it was denser at these earlier epochs, so that
interactions and collisions between self-gravitating objects were
frequent. This certainly explains why galaxies observed at large
distances seem rather disturbed and do not fit the Hubble morphological
classification (van den Bergh 1998).

There are typically about 10\textsuperscript{8} -- 10\textsuperscript{12
}\ stars in a galaxy. Once formed from the collapse of a cloud of gas
and dust, stars evolve by themselves depending on their initial mass
and chemical composition. The environment has very little effect on
them, they essentially never collide, die, nor disappear. Only the most
massive ones explode as supernovae, injecting into the interstellar
medium some of their gas that is enriched in heavy atomic elements
({\textquotedblleft}metallic{\textquotedblright}, i.e. heavier than
Beryllium and Lithium) and leaving a dense remnant that can be a
neutron star or a small black hole. 

Galaxies generally have a lot of gas and dust which properties evolve
because of stellar radiation and density perturbations. The chemistry
of atoms, molecules and grains is local and very complex, but the
orbital motions rather rapidly homogenize somehow the chemical
properties within a galaxy. In particular, \ the collapse of molecular
clouds that leads to the formation of stars generally happens \ more or
less simultaneously over a significant fraction of the volume of the
galaxy. This phenomenon relates global characteristics of galaxies with
small scales processes, but it also smears out traces of past localized
events.

The properties of the three \index{fundamental constituents}fundamental
constituents (stars, gas and dust) fully describe a given galaxy,
provided we have access to them at all places in the three dimensions.
This implies that the full \index{description}description of a galaxy
is already complex. Moreover, any gravitational perturbation, either
caused by external passing-by galaxies or by internal interactions,
modifies the shape of the galaxy, and possibly properties of gas and
dust. Even internal events, like the explosion of many supernovae at
nearly the same time or density waves, may have a strong influence.
Since there are so many components in a single galaxy, since their
mutual interactions through radiation, shock waves, gravitational
perturbations, chemical and physical processes and so on, are
essentially non-linear and chaotic, the evolution of a galaxy truly
belongs to the \index{complex}complex sciences.

Nowadays, observations are becoming very detailed and numerous.
Systematic surveys feed databases comprising millions of galaxies for
which we have images and spectra. Big telescopes and very sensitive
detectors allow us to observe objects from the distant past (galaxies
at high redshifts). This is somewhat reminiscent of paleontology, and
like evolutionary biologists, astronomers want to understand the
relationships between distant and nearby galaxies, like our own.
Strangely enough, the Hubble classification, born from the Hubble
diagram, is still frequently used as a support to describe galaxy
evolution, even though it ignores all observables except morphology
(i.e. \textcolor{black}{Hernandez and }\textcolor{black}{Cervantes-Sodi
2006, Cecil and Rose 2007}). Indeed, the Hubble classification is very
successful not only because it is simple but also because the global
shape of a galaxy grossly summarizes several physical properties. In
particular, star formation is more efficient in disks and it seems that
spiral galaxies have also more gas than ellipticals. Obviously, the
distribution of the stellar orbits is related to the history of a
galaxy. Numerical simulations are easier to make with stars and
gravitation only, the gas and dust components requiring more
complicated equations and much more computer power. Thus, many
simulations have shown precisely how the structures of galaxies
transform themselves during interactions or merging (Bournaud et al.
2005). For instance, ellipticals have been shown to result from the
merging of two galaxies of similar masses. However, the inclusion of
gas in the simulations now leads to other possible formation scenarii
(Bournaud et al. 2007, Ocvirk et al. 2008).

Other \index{classification}classifications of galaxies are somehow
inherited from this morphological Hubble classification and are most
often dictated by the instrument used to make the observations of the
sample. Correlation plots are used to make crude classification in a
very few number of categories (often 2 like blue and red, high and low
intensity for a given wavelength, more or less metallic, ...). This is
most of the time sufficient for some physical modelling, but it cannot
describe the huge diversity of galaxies across the Universe and their
now recognized complex history in a very objective and synthetic way.

The understanding of galaxy \index{formation}formation and
\index{evolution}evolution, which we prefer to call galaxy
\index{diversification}diversification, is a major challenge of
contemporary \index{astrophysics}astrophysics. Abandoning the
one-parameter classification approach and using all available
descriptors means taking a methodological step equivalent to the one
biologists took after Adanson and Jussieu in the 18th century. Today
the tools do exist, and ordination methods, essentially the one of
Principal Component Analysis, are being used more often, mainly to
automatically separate stars and quasars from galaxies on large images
of the sky (e.g. Cabanac et al. 2002). A few attempts to apply
clustering methods have been made recently, still with little success
in identifying \ new classes that convince the astronomical community
(e.g. Chattopadhyay \& Chattopadhyay 2006). From our point of view,
there are two difficulties here. The first one is that the PCA
components are non-physical, hence very difficult to interpret and
model in the way astrophysicists are used to. The second one, more
important in our point of view, is that evolution, an unavoidable fact,
is not at all taken into account. By mixing together objects at
different stages of evolution, any physical significance of a
classification is undoubtedly lost. This was precisely our motivation
for developing astrocladistics in 2001 (Fraix-Burnet et al 2006a,
2006b, 2006c).

\section[2. The astrocladistics project]{ 2. The
astrocladistics project}
\index{astrocladistics}\section[2.1 A phylogenetic framework for the
galaxies]{2.1 A phylogenetic framework for the galaxies}
\index{phylogenetic}

As we have seen in the Introduction (Sect. 1), the baryonic matter is
immersed in the gravitational field that is shaped by the dominating
dark matter. Even though its nature is totally unknown, the dark matter
is supposedly affected only by gravitation. This is a relatively simple
physics and it enabled the first numerical simulations of the
cosmological evolution of the gravitational field in the expanding
Universe from the observed primordial fluctuations up to the present
time. Since matter attracts matter, tiny overdensities, called halos,
progressively grow in mass and size by merging together. This is the
\index{hierarchical}hierarchical model of formation of the large
structures in the Universe.

The evolution of dark matter halos is generally represented by a
{\textquotedblleft}merger tree{\textquotedblright}, a typical one being
shown on Fig. 1a (Stewart et al. 2008). Many small halos at the top
(redshift of about 7 equivalent to about 13 Gyr ago) of the tree merge
while time goes downward to yield large dark matter halos observed in
galaxy clusters. These merger trees indeed represent the genealogy of a
single halo. A schematic representation these trees is often used, but
they tend to suggest that small halos disappear with time. This is not
true, and the tree on Fig. 1a clearly shows that some of them
{\textquotedblleft}survive{\textquotedblright}. Hence, at any epoch in
the Universe, halos of different sizes with different merging histories
coexist.

From the astrocladistics point of view, the
\index{hierarchical}hierarchical evolution of the dark matter halos is
better represented on a phylogenetic tree, where mass (or size) is the
only criterion for diversity (Fig. 1b). This tree describes the
evolution of the environment in which galaxies form and evolve. \

Galaxies are made up of baryonic matter which is sensitive to
gravitation, but also to electromagnetic, weak and strong interactions,
as well as radiation and thermodynamics. Its physics is thus very
complex and there are many processes that can strongly affect its
gravitational behaviour. Probably because of the first cosmological
numerical simulations that were able to take the sole gravitation into
account, galaxies are generally thought of as being at the center of
dark matter halos. As a consequence, the hierarchical formation
scenario seems to apply naturally to galaxies, and has been very
popular up to now (e.g. Baugh 2006, Avila-Reese 2007).

However, the old formation scenario for galaxies, called the monolithic
model, gains new consideration \ thanks to new numerical simulations
and observations. This model indeed describes the collapse of a cloud
of gas, and includes all the necessary physics which is not included in
the pure hierarchical model. Merging of galaxies does exist, but it
cannot be of the same nature as for the dark matter since the
gravitational energy can be transformed into heat, shocks or radiation.
Consequently, the diversification of galaxies cannot follow the same
simple hierarchical scenario as that of the dark matter represented on
Fig. 1. 

\begin{figure}[t]
\begin{center}
\includegraphics[width=11.666cm,height=11.261cm]{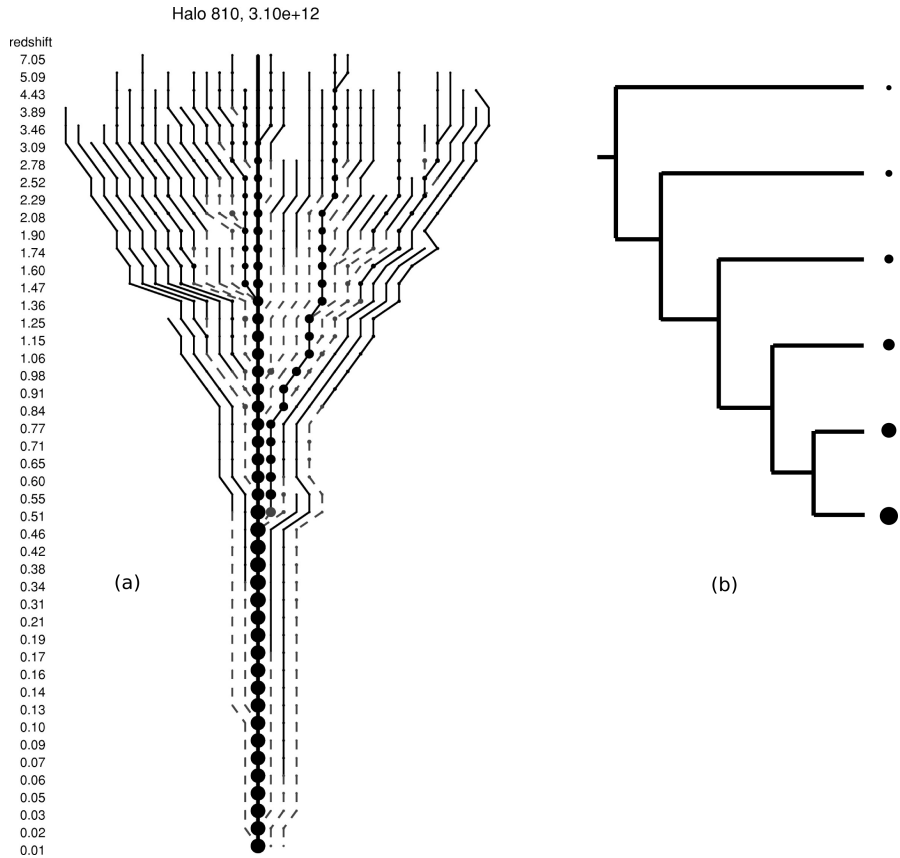}
\caption{(a) A {\textquotedblleft}typical{\textquotedblright} merger
tree for an individual dark matter halo with mass ${\simeq}$
10\textsuperscript{12.5} solar masses at \ present time (redshift=0).
Time progresses downward, with the redshift printed on the left
hand side. The bold, vertical line at the center corresponds to the main
progenitor, with filled circles proportional to the radius of each
halo. The minimum mass halo shown in this diagram has m =
10\textsuperscript{9.9} solar masses. Solid and dashed lines and
circles correspond to isolated field halos, or subhalos,
respectively. The dashed lines that do not merge with main progenitor
represent surviving subhalos
at z = 0. Such trees are indeed
genealogical trees for individual halos. Taken from Stewart et al 2008 and reproduced by permission of the AAS. (b) The
corresponding
phylogenetic tree as seen by astrocladistics. There is no time scale in
this representation. Mass or size is the sole diversification criterion
and increases downward. Each bifurcating node corresponds to a merger
event and indicates the first occurrence in the Universe of a halo of a
given mass. The chronology thus goes from left to right.}
\label{fig1}
\end{center}
\end{figure}

Astrocladistics makes it clear that the formation and the evolution of
galaxies, governed by a complex physics, occur in an
\index{environment}environment whose evolution, distinct and largely
independent, is governed by the sole gravitation. Galaxies interact
between each other while they move in the gravitational field shaped by
the dark matter. To a first approximation, galaxies can be good tracers
of the dark matter, but as recent observations show, the two can
sometimes be well disconnected.

It is still difficult to tell what the very first objects were. Since
the Universe was essentially homogeneous at the recombination epoch,
the primordial gas was certainly not very clumpy. With time, with the
growth of the tiny initial fluctuations in dark matter halos, gas
density began to increase in some places. When these seeds became
gravitationally bound, we could call them galaxies. It is probable that
the first stars formed very rapidly during this process. These very
first galaxies were probably already quite diverse in mass content
(gas/stars) but very similar otherwise. They also subsequently merged
rapidly, interacted strongly with their neighbours, forming new and
still more diverse objects. The expansion of the Universe definitively
separated large regions that make what we call superclusters and
clusters of galaxies. These can be seen as islands in which the
corresponding populations of galaxies have evolved on their own,
possibly creating some evolutionary divergence.

However, the physics is the same everywhere (cosmological principle),
the environment is purely gravitational and differs from place to place
only by its shape in the curved space-time. We thus expect a lot of
similarities between different
{\textquotedblleft}populations{\textquotedblright} of galaxies due to
convergent and parallel evolutions. Unfortunately, our quest to
understanding diversification is somewhat complicated by the finite
velocity of light. We have not access to the entire Universe, hence to
the entire galaxy diverstiy, that is contemporary to us because when
galaxies are a bit far away, we see them as they where when the light
was emitted. We are like paleontologists, except that we do not have
access to the entire present diversity. Nonetheless recent observations
seem to suggest that galaxy evolution has been very gentle in the last
1Gyr or so.

\section[2.2 Transmission with modification among galaxies]{
2.2 Transmission with modification among galaxies}

\begin{figure}[t]
\begin{center}
\includegraphics[width=7.578cm,height=3.03cm]{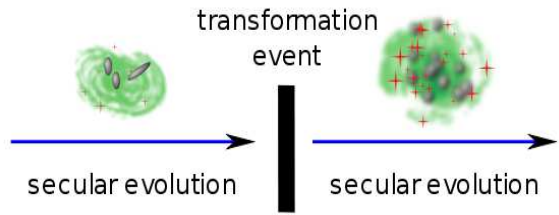}
\caption{A schematic view depicting the transmission with modification
process in galaxy diversification. \ For any transformation event, gas,
dust and stars are transmitted to the new object generally with some
modification of their properties. Secular evolution is defined as the
evolution of a totally isolated galaxy. It thus occurs all the time,
and even if it is not violent in general, it can sometimes lead to a
significant transformation.}
\label{fig2}
\end{center}
\end{figure}

Galaxy formation is now acknowledged to occur continuously during the
Hubble time (Hoopes et al. 2007). In extragalactic astrophysics,
evolution and formation are generally not clearly distinguished.
Indeed, galaxy evolution is more like galaxy transformation, that leads
to the formation of new kinds of objects. Formation often seems to be
employed as if big spiral and elliptical galaxies are at the end of
evolution. To our point of view, {\textquotedblleft}galaxy formation
and evolution{\textquotedblright} can be more precisely called galaxy
diversification, in which evolution means transformation and formation
refers to diversity.

There are 5 \index{transformation}transformation processes for galaxies:
assembling, secular evolution, interaction, accretion/merging,
ejection/sweeping. In all these cases, the fundamental material of a
galaxy, stars, gas and dust, is transmitted to the new object, somewhat
modified. These modifications are mainly kinematical because the
gravitational perturbations change the orbits, hence the distribution
of the objects. The stars are otherwise unaffected, except that new
ones can be formed from the perturbed gas and dust. These latter two
constituents can also be modified in density, temperature and
composition.

Schematically, galaxy diversification can be depicted like in Fig. 2. At
a given time, a galaxy undergoes a transformation event to form a new
object made with the same constituents but modified. Between two such
events, stars get older and essentially cooler (redder). This is part
of the secular evolution, which is the evolution of a totally isolated
galaxy. Secular evolution is a transformation process driven by
internal phenomena (stellar ageing, interaction between the three
constituents, density waves, shock waves due to supernovae,
instabilities, ...) which is generally slow and non-violent. However,
after a certain time, the galaxy can be so different that it can be
described as a new object and even belong to a different class of
objects. Like for living species, the modifications are gradual and at
some level of details, the splitting in different species is somewhat
arbitrary. Thus, secular evolution can also be a transformation event
participating to the diversification of galaxies.

Figure 2 shows that it is possible to see the transformation of galaxies
as a \index{transmission with modification}transmission with
modification process. It is not exactly a darwinian process in the
sense that there is no natural selection (even though big galaxies
survive more easily and tend to swallow small ones!). Also, there is no
duplication, but each transformation process is a replication
mechanism, because the new object has inherited all constituents and
most of their properties. Since the transformation events are always
random, this is equivalent to a replication with
{\textquotedblleft}spontaneous mutations{\textquotedblright}. In
addition, the environment, which is gravitational, has a strong
influence on the occurrence and properties of all the galaxy
transformation processes.

Galaxy diversification is thus characterized by replication, randomness
and environment. It follows \ a branching pattern can be expected. This
is why \index{cladistics}cladistics should be applicable to this
problem. However, it might not be so simple since mergers are probably
frequent at least near the beginning of the Universe, so that
reticulation ({\textquotedblleft}hybridation{\textquotedblright} or
horizontal transfer) could be present. Cladistic analyses of several
samples of galaxies can investigate this. As shown in Fraix-Burnet et
al. (2006a, 2006b) and confirmed by all the \ robust cladograms we have
obtained so far, \index{reticulation}reticulation does not seem to be
dominant. Yet, this conclusion must be seen as preliminary and more
investigation should be done, particularly with very distant objects.

\section[3. Applying cladistics to galaxies]{ 3. Applying
cladistics to galaxies}

Astrophysics is an observational science, not an experimental one. It is
not possible to weigh a galaxy or to return it to see what is behind.
We simply collect photons, at all wavelengths (from gamma-rays and
X-rays to the radio domain, with the ultraviolet, optical, infrared,
far-infrared and sub-mm in between), that are emitted by every
component at every place in a galaxy. The level of details is limited
by the faintness and the small apparent projected size of the objects,
particularly those at very high distances from us. Both improve with
the diameter of the photon collector (the mirror of the telescope) and
with the sensitivity of the detectors. Requirements for the
astronomical observations are a strong thrust for technology research
and development, and recent progresses in this domain have changed the
scale in the amount of data concerning galaxies. But still, each
picture element ({\textquotedblleft}pixel{\textquotedblright}, i.e. the
smallest detail) is a mixture of the light emission from many stars and
gas/dust clouds situated along the line of sight and across a region
which size depends on the distance of the galaxy.

Basically, observations result in the recording of spectra. Apart from
the structure, given by imagery or relative spatial positioning,
spectra contain all the physical and chemical information that is
possible to get from remote objects. Systematic surveys of the sky now
provide spectra for millions of galaxies. However, in addition to the
limitations mentioned above, there are limits coming from the spectral
resolution of the detectors and from Doppler effects due to the
differential motions of the galaxy components that enlarge and blur
absorption and emission lines. As a consequence, physical and chemical
conditions at each part of a galaxy are always partial and averaged.

Measuring characteristics from spectra is also not sufficient to derive
physical or chemical quantities, some combinations must be used like
ratios of several emission lines. Moreover, models must most often be
used to disentangle the complexity of the emitting region and to
translate spectral information into physical quantities. Unfortunately,
these models necessarily introduce some subjectivity. Thus, truly
intrinsic descriptors of galaxies are not easy to obtain, and this is
certainly a specific difficulty of astrocladistics.

Anyhow, descriptors in extragalactic astrophysics are always
\index{continuous variables}continuous variables since, apart from
spectral information, the other information we can measure is the
dimension (size and position of individual components within the
galaxy). The only exception is the morphology. It is traditionally
based on the Hubble classification represented by the Hubble diagram,
most often with the use of a discrete scale distinguishing different
kinds of elliptical and spiral galaxies. However, in our astrocladistic
project, we have chosen not to use this parameter for several reasons.
The first one is that this is a very crude way of discretizing a shape
parameter because it is essentially made by eye. The second reason is
that this discretization is not homogeneous in the sense that
differences between spirals are given the same weight as between
ellipticals, while they are not of the same nature (shape and number of
the spiral arms which are density waves, not physical structures, for
the first ones, and estimation of roundness for the latter ones ; de
Vaucouleurs 1994). The third reason is that it is not clear at what
level this descriptor can be informative regarding evolution. Finally,
we know that the Hubble classification is correlated with quite a few
global properties of galaxies (kinematics, amount of gas and star
formation most notably), so it is certainly redundant with other
descriptors, quantitative and objective, coming from the spectrum.

For such an exploratory project like astrocladistics, choices have to be
made right at the beginning to focalize the research. Up to now, we
have chosen to discretize all observables into 10 or 30 bins, depending
on the error bars estimated by the observers and on the distribution of
values among the sample. This allows us to use parsimony as the
optimisation criterion, which seems to be the simplest strategy to
implement for a cladistic analysis. These first choices already enabled
us to obtain very positive results. Other paths can now be explored.

Last but not least, we do not have a multivariate classification of
galaxies at our disposal. We have to assume that each galaxy represents
a {\textquotedblleft}species{\textquotedblright} that will have to be
defined later on. It is consequently \ difficult to define an outgroup
that is necessary to root the tree and help the interpretation of the
evolutionary scenario. Most generally, we root the trees with the
object that is the less metallic, that is in which the relative
abundance of heavy atomic elements is the lowest (heavy elements are
only produced in stars, and cannot be primordial). But we are aware
that this choice is made on one character only.

\section[4. The first extragalactic trees]{ 4. The first
extragalactic trees}
The first application of astrocladistics was performed on a sample of
galaxies issued from a cosmology simulation. This work used 50 objects
and 50 characters (observables), corresponding to 5
{\textquotedblleft}lineages{\textquotedblright} sampled at 10 epochs.
In such simulations, each entity, called a galaxy, that appears, is
given a number. Only accretion or merging are considered, and a new
number is attributed each time such an event occurs. It is then easy to
follow the tree of transformations of an initial galaxy into its
different descendants. The simulations calculate the radiation from
galaxies and transpose it into the same observables as obtained from
real telescopes. This work was intended to illustrate the concepts and
practical methodology of astrocladistics, and to show how the correct
{\textasciigrave}{\textasciigrave}genealogy{\textquotesingle}{\textquotesingle}
can be reconstructed (Fraix-Burnet et al. 2006a, 2006b) from usual
observables.

\begin{figure}[t]
\begin{center}
\includegraphics[width=12.268cm,height=12.097cm]{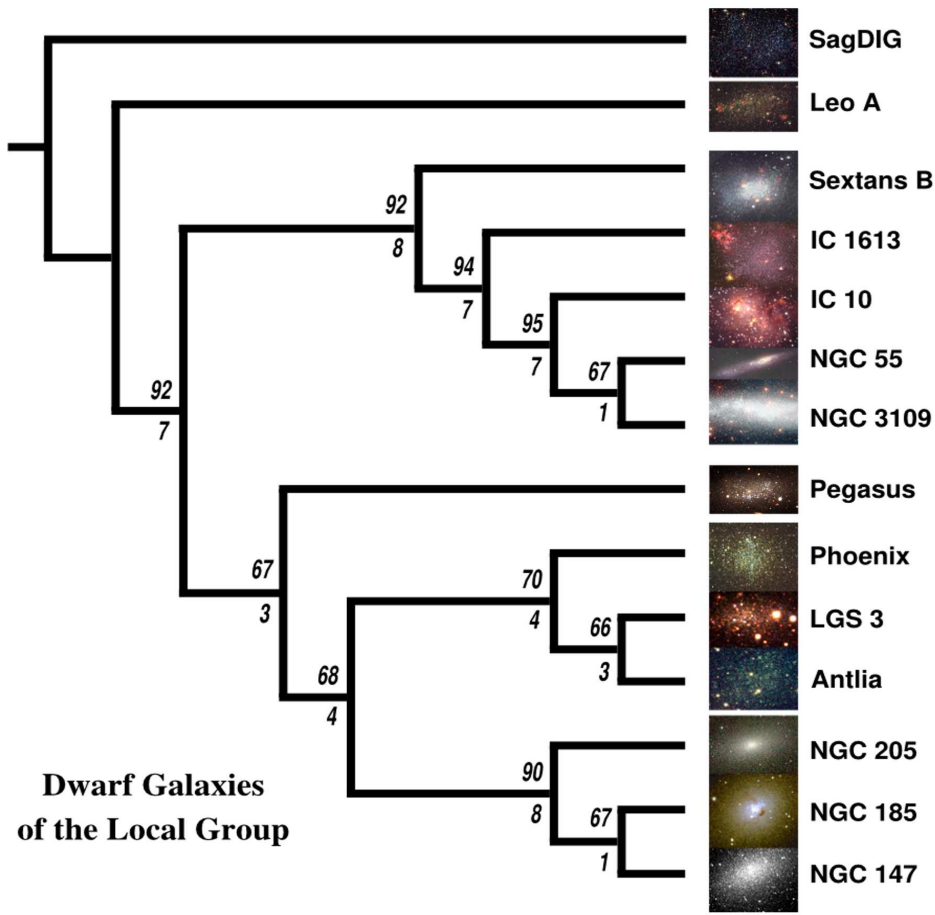}
\caption{Cladogram of 14 Dwarf galaxies of the Local Group obtained with
24 characters (observables and derived quantities). Bootstrap values
(above) and Decay indices (below) are indicated for each node. The
outgroup (SagDig) has been chosen because it contains the lowest amount
of metallic material, suggesting that it is made up of more primordial
material (see Fraix-Burnet et al 2006c for details on this work). Image credits: NGC~55 : David Malin; LGS~3 and
Pegasus~DIG: Deidre A. Hunter; Antlia: Mike Irwin; NGC~185: David M. Delgado; NGC~147: Walter Nowotny; Sag~Dig: Hubble
Heritage Team (AURA / STScI), Y. Momany (U. Padua) et al., ESA, NASA; Leo~A, Sextans~B, IC~1613 and IC~10: Corradi,
R.L.M. et al., 2003, ING Newsletter No.~7, p. 11; NGC~3109: NASA/ STScI; Phoenix: Knut Olsen (CTIO) \& Phillip Massey
(Lowell Observatory), (NOAO / CTIO / KPNO); NGC~205: Atlas Image [or Atlas Image mosaic] courtesy of
2MASS/UMass/IPAC-Caltech/NASA/NSF.}
\label{fig3}
\end{center}
\end{figure}

\begin{figure}[!ht]
\begin{center}
\includegraphics[width=15.889cm,height=21.426cm]{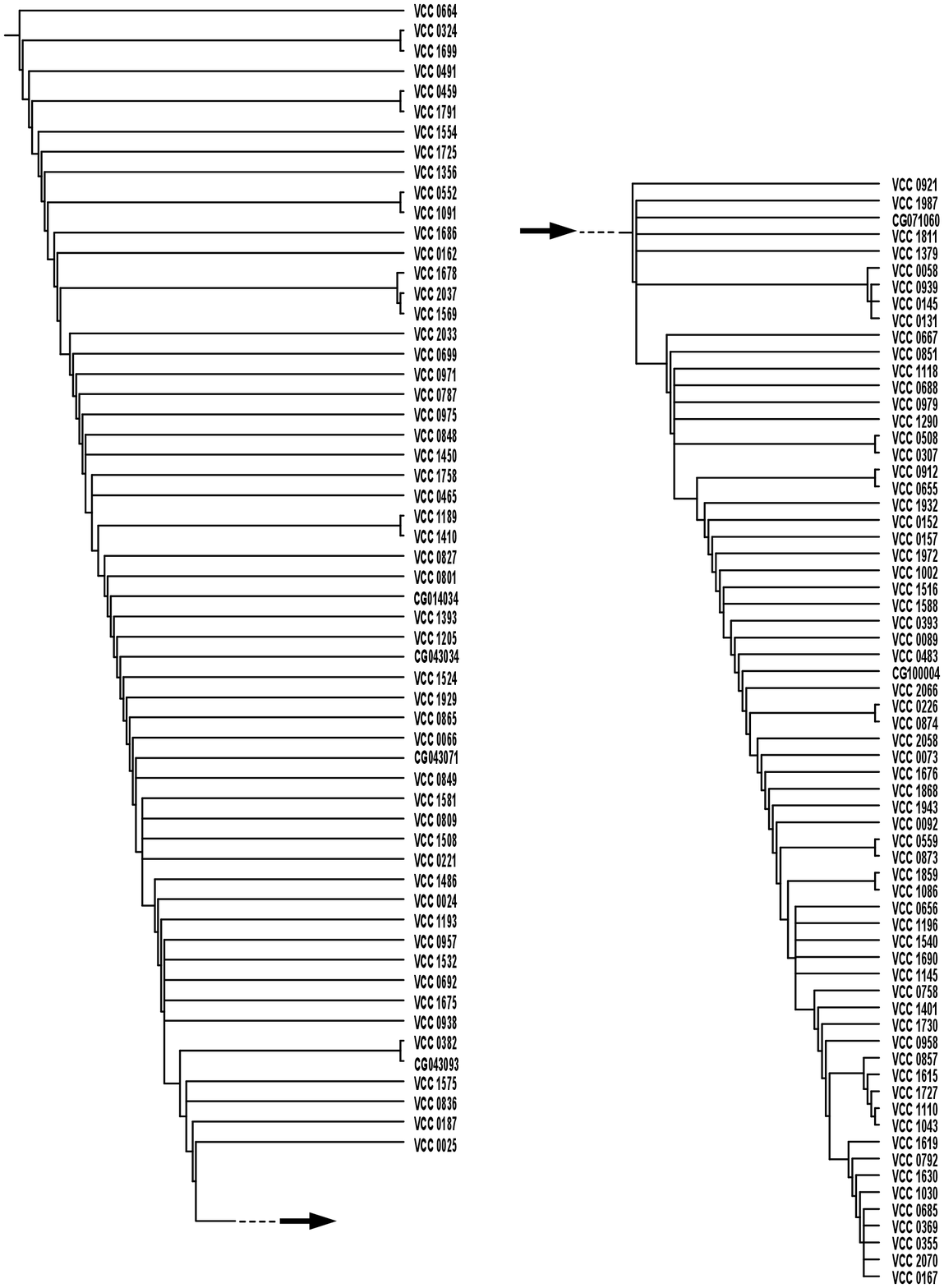}
\caption{Cladogram for 123 galaxies of the Virgo cluster with 41
observables. Note the regularity of the tree indicating the cosmic
evolution (star ageing) which is highly redundant. }
\label{fig4}
\end{center}
\end{figure}

More importantly, a small sample of real galaxies (Dwarf Galaxies of the
Local Group, 36 objects, 24 characters) have also successfully been
analysed, providing the first evolutionary galaxy tree ever established
(Fraix-Burnet et al. 2006c). These dwarf galaxies are very numerous
throughout the Universe. Most of them are satellites of big galaxies
like our own. They supposedly have been the first galaxies at the
beginning of the Universe and are thus believed to be the building
blocks that merged to make bigger and bigger galaxies. A fully resolved
tree obtained on a subset is represented in Fig. 3. Bootstrap and decay
values indicated at the nodes show that it is very robust. The
groupings are consistent with our current knowledge of these particular
objects, with a more refined classification. This tree clearly implies
that the nature of the ancestors are of irregular shape. The other
type, the spheroidals, are consequently more evolved. This question is
debated in the astrophysical community and is naturally addressed by
astrocladistics in a multivariate way. It is important to note that the
characters used in this analysis are a mixture of observables (like
colours, fluxes in different spectral lines and composition ratios) and
derived quantities (like masses, maximum rotational velocities,
velocity dispersion, star formation rate). Such kind of information is
more difficult to obtain with other more distant galaxies.

We subsequently performed a cladistic study of 222 galaxies of the Virgo
cluster using 41 characters that were all observables. Because of too
many unknown parameters, we were able to build a robust tree for 123
objects only, and used it to construct a supertree. Figure 4 shows the
cladogram with 123 galaxies. It took us some time to understand the
astrophysical consequence of the regularity of the (largely unbalanced)
tree that is also found on the supertree. In fact, this is due to the
sole stellar evolution which is hidden in nearly all of the
observables. It is consequently highly redundant and dominant in this
multivariate analysis. The astrophysical outcome is thus slightly
disappointing. We obtained similar results on other samples: poor gas
galaxies of the field (comprising 227 objects and 37 characters),
galaxies from a big survey (500 objects with 60 characters analysed so
far).

From the astrocladistic point of view, all these studies shows that our
approach is successful in constructing cladograms of galaxies in
several different types of samples. Nevertheless, the regular trees
found when using solely observables reveal that stellar evolution is
universally present and acts as a parallel evolution. It does not bring
much information on galaxy diversity, and disturbs the multivariate
analyses. We call it {\textquotedblleft}\index{cosmic evolution}cosmic
evolution{\textquotedblright} because it is universal and does not lead
to a sufficient transformation of the galaxy, like secular evolution in
general or the other transformation processes. Even in the most violent
cases, a given stellar population continues to age unperturbed and its
radiation properties are unaffected unless the population itself is
changed for instance with the addition of newborn stars. Somehow, we
could think the cosmic evolution as a kind of ontogeny for galaxies. We
looked for a way to remove it from the observables because it obviously
hampers the search for
{\textquotedblleft}speciation{\textquotedblright}. We used various
synthetic stellar populations, modelling artificial lineages with
specific properties. We restricted ourselves to small samples (30
objects). Our idea was to scrutinize each observable in a very well
controlled sample. By trials and errors, we found some combinations of
several observables \ that make the corrected characters to remain
essentially constant between transformation events (Fraix-Burnet et al.
in prep). The physical interpretation of these corrected characters is
still unclear, but the important result is that they lead to physically
consistent clades for these simple synthetic stellar populations.
Application to real objects is under progress. 

Globular clusters are self-gravitating ensembles of stars without dust
or gas. In a sense, they are simple galaxies, somehow like our
synthetic stellar populations. Taking advantage of our experience
gained so far, we performed a cladistic analysis of 54 globular
clusters of our Galaxy by choosing 4 properties. Three of them are
intrinsic and comes from their conditions of formation. They are not
affected by cosmic evolution. The other one is the colour and measures
the age of the stars, it is thus a direct indicator of the cosmic
evolution. We gave it a weight half of the other ones, and effectively
rank the clusters chronologically within the lineages or groupings that
are defined by the other three characters. Our result tree gathers the
objects in a new classification that brings a very convincing scenario
for the formation of these objects and consequently for some episodes
in the building up of our own galaxy. This result is the first
spectacular outcome of astrocladistics and will hopefully convince more
astronomers to invest in this novel approach (Fraix-Burnet, Davoust \& Charbonnel 2009).

\section[5. Some open questions]{ 5. Some open questions}
It is now clear that cladistics can be applied and be useful to the
study of galaxy diversification. Many difficulties, conceptual and
practical, have been solved, but many remain in the exploration of this
large research field. To begin with, it is well possible that
probabilistic methods, instead of parsimony, might be better suited for
the quantitative nature of the variables and the evolution of galaxies.
In this respect, the use of the continuous data without discretization
will have to be investigated seriously, even with parsimony (Wiens
2001, MacLeod 2002, Goloboff et al. 2006, Gonz\`alez-Ros\'e et al.
2008). \

There are difficulties that seem to be intrinsic to astrophysics. Most
notably, we have millions of objects, but a few tens of descriptors. Of
course the situation will improve with time, in particular with
integral-field spectroscopy (spectra of detailed regions in the galaxy,
e.g. Ensellem et al. 2007). Spectra might not be currently employed at
their full capacity of description. But it is not clear whether they
would lead to the discrimination of hundreds of classes. Perhaps this
is an erroneous target, perhaps galaxies cannot be classified with such
a refinement. But this is already a matter of using multivariate
clustering methods and interpreting their results usefully. We are
convinced that improvements can be made here. In any case, cladistics
is supposed to identify clades, that are evolutionary groups, whereas
the concept of {\textquotedblleft}species{\textquotedblright} is not
defined at all in astrophysics, and we have even not converged toward
groupings based on multivariate analyses.

This is probably more than a conceptual question because clustering with
continuous data and large intraspecific variance is a very complicated
problem in itself. Sophisticated statistical tools must be used, but
the question of characterizing the groups in this context is not very
clear to us at this time. This probably requires a different culture
that is not yet present in astrophysics. Incidentally, this also points
to the relevance of discretization we have used so far particularly
when measurement uncertainties are sometimes quite large. Our results
definitively show that we are not wrong, but it might be possible to do
better.

The possible predominance of reticulation is a concern for cladistics in
general, and we foresee to investigate this question very soon. In some
ways, galaxy diversification has similarities with bacteria
diversification for which reticulation methods have been considerably
developed recently. We intend to take advantage of these advances. It
would already be interesting to detect some reticulation in the samples
we have studied so far. The equivalent of hybridation or horizontal
transfer certainly comes through the merging of two galaxies. Since the
Universe is in expansion, object density was higher at earlier times,
and interactions and merging were more common. As a consequence,
reticulation could be a problem for distant galaxies, that is when the
Universe was young.

A very important step for astrocladistics is our assessment of the
influence of the cosmic evolution and our ability to correct for it or
avoid it in the character selection. But there are probably other kinds
of parallel or convergent evolutions, and reversals as well. Hence, we
are now concentrating our efforts on finding the most pertinent
descriptors. For this we must detail all the possible quantities we can
get from the observations, to see how they behave during the different
transformation processes and whether they can be reliable tracers of
past events from a cladistic point of view. Biologists pretty well know
that this is a crucial work, apparently without an end.

\section[6. Conclusion]{ 6. Conclusion}
The astrocladistic project has now reached a solid base as far as
concepts and tools are concerned. \ First results have shown the
usefulness of this approach, and unavoidable difficulties are being
solved one after one. This triggers the hope of being able to map
galaxy evolution from the very first objects of the Universe, some 14
Gyr ago, but there is still a long way to go. Astrocladistics opens a
new way to analyse galaxy evolution and a path towards a new
systematics of galaxies.

Extragalactic astrophysics is currently undergoing a scaling revolution.
The number of objects and the amount of data it has to cope with
already impose new statistical tools to be used, and this will
certainly get worse with still bigger telescopes and more sensitive
detectors. Less than a century after their discoveries, galaxies come
by millions in our databases. This is a cultural revolution, in
particular because the complexity of these objects and their evolution
become more obvious.

Astrocladistics is fascinating in combining a multivariate statistical
philosophy with evolutionary concepts. It is fascinating as an
interdisciplinary research field. But it is also quite a difficult
approach, not intuitive at all, and does not fit well in the usual
methods of physicists. This is a huge research field, very exploratory,
with many possible paths to follow. We still lack spectacular
astrophysical outcomes, but most astrophysicists are very enthusiastic,
because they agree that there is an obvious need for multivariate and
evolutionary methods to study galaxy diversification. 

\section[References]{ References}

\begin{description}
 \item[]
Avila-Reese V. (2007) Understanding Galaxy Formation and Evolution.
Ap\&SS Proceedings, Solar, stellar and galactic
connections between particle physics and astrophysics, Carrami{\~n}ana A, Guzm\'an F S, Matos T (eds),
Springer (2007), 115-165. \url{http://arxiv.org/abs/astro-ph/0605212}
 \item[]
 Baugh C.M. (2006) A primer on hierarchical galaxy
formation: the semi-analytical approach. Reports on
Progress in Physics, Volume 69, Issue 12, 3101-3156.
\url{http://arxiv.org/abs/astro-ph/0610031}
 \item[]
Bournaud F., Jog C.J., Combes F. (2005) Galaxy mergers with various mass
ratios: Properties of remnants. Astronomy and Astrophysics 437, 69-85.
\url{http://arxiv.org/abs/astro-ph/0503189}
 \item[]
Bournaud F., Jog C.J., Combes F. (2007) Multiple minor mergers:
formation of elliptical galaxies and constraints for the growth of
spiral disks. Astronomy and Astrophysics 476, 1179-1190.
\url{http://arxiv.org/abs/0709.3439}
 \item[]
Cabanac RA, de Lapparent V, Hickson P (2002) Classification and
redshift estimation by principal component analysis. Astronomy \&
Astrophysics \ 389, 1090--1116. \url{http://arxiv.org/abs/astro-ph/0206062}
 \item[]
Cecil G., Rose J.A. (2007) Constraints on Galaxy Structure and Evolution
from the Light of Nearby Systems. Rep. Prog. Phys. 70, 1177-1258.
\url{http://arxiv.org/abs/0706.1332v2}
 \item[]
Chattopadhyay T., Chattopadhyay A.K. (2006) Objective classification of
spiral galaxies having extended rotation curves beyond the optical
radius. The Astronomical Journal 131, 2452--2468.
 \item[]
 de Vaucouleurs G. (1994) Global Physical Parameters of Galaxies. In:
Quantifying Galaxy Morphology at High Redshift. Space Telescope Science
Institute (Baltimore, USA), April 27-29 1994. \\
\url{http://www.stsci.edu/institute/conference/galaxy-morphology/program2.html}.
 \item[]
Emsellem E., Cappellari M., Krajnovi\'c D., van de Ven G. , Bacon R.,
Bureau M., \ Davies R.L., de Zeeuw P.T., Falc\'on-Barroso J.,
Kuntschner H., McDermid R., Peletier R.F., Sarzi M. (2007) The SAURON
project -- IX. A kinematic classification for early-type galaxies.
Mon. Not. R. Astron. Soc.
379, 401-417.
\url{http://arxiv.org/abs/astro-ph/0703531v3}
 \item[]
 Fraix-Burnet D, Choler P, Douzery E, Verhamme A (2006a) Astrocladistics:
a phylogenetic analysis of galaxy evolution. I. Character evolutions
and galaxy histories. Journal of Classification 23, 31-56. \\
\url{http://arxiv.org/abs/astro-ph/0602581}
 \item[]
Fraix-Burnet D, Douzery E, Choler P, Verhamme A (2006b) Astrocladistics:
a phylogenetic analysis of galaxy evolution. II. Formation and
diversification of galaxies. Journal of Classification 23, 57-78. \\
\url{http://arxiv.org/abs/astro-ph/0602580}
 \item[]
Fraix-Burnet D, Choler P, Douzery E (2006c) Towards a
Phylogenetic Analysis of Galaxy Evolution : a Case Study with the Dwarf
Galaxies of the Local
Group, Astronomy \&
Astrophysics 455, 845-851.\\ 
\url{http://arxiv.org/abs/astro-ph/0605221}
 \item[]
Fraix-Burnet, D., Davoust, E., Charbonnel, C. (2009) The environment of formation as a second parameter for globular cluster classification.  Monthly Notices of the Royal Astronomical Society, in
press. \\
\url{http://fr.arxiv.org/abs/0906.3458}
 \item[]
Goloboff P.A., Mattoni C.I., Quinteros A.S. (2006) Continuous
characters analyzed as such. Cladistics 22, 589--601.
 \item[]
Gonz\`alez-Jos\'e R., Escapa I., Neves W.A.,
C\'uneo R., Pucciarelli H.M.
(2008) Cladistic analysis of continuous modularized traits provides
phylogenetic signals in Homo evolution. Nature 453, 775-779.
 \item[]
 Hernandez X, Cervantes-Sodi B (2006) A dimensional
study of disk galaxies.
11th Latin-AmericanRegional IAU
Meeting 2005, RevMexAA (Serie de Conferencias), 26, 97--100. \url{http://arxiv.org/abs/astro-ph/060225}
 \item[]
Hinshaw G., Weiland J.L., Hill R.S., Odegard N., Larson D., Bennett
C.L., Dunkley J., Gold B., Greason M.R., Jarosik N., Komatsu E., Nolta
M.R., Page L., Spergel D.N., Wollack E., Halpern M., Kogut A., Limon
M., Meyer S.S., Tucker G.S., Wright E.L. (2009) Five-Year Wilkinson
Microwave Anisotropy Probe (WMAP) Observations: Data Processing, Sky
Maps, and Basic Results. Astrophysical Journal Supplements180:225-245.
\url{http://arxiv.org/abs/0803.0732}
 \item[]
Hoopes C.G., Heckman
T.M., Salim S. et al. (2007) The Diverse Properties of the Most
Ultraviolet-Luminous Galaxies Discovered by GALEX. Astroph. J. Suppl.
Series 173, 441--456. \\
\url{http://arxiv.org/abs/astro-ph/0609415}
 \item[]
Kashlinsky A., Arendt R.G., Mather J., Moseley S.H. (2005) Tracing the
first stars with fluctuations of the cosmic infrared background.
\textcolor{black}{Nature 438}\textcolor{black}{, 45-50.
doi:10.1038/nature04143}
 \item[]
MacLeod N. (2002). Phylogenetic signals in morphometric data, pp.
100-138. In: MacLeod N., and Forey, P.L. (eds.), Morphology, Shape, and
Phylogeny. Taylor \& Francis, London.
 \item[]
Ocvirk P., Pichon C., Teyssier R. (2008) Bimodal gas accretion in the
Horizon-MareNostrum galaxy formation simulation.
Mon. Not. R. Astron. Soc. 390, 1326-1338.
 \item[]
Stewart K.R., Bullock J.S., Wechsler R.H., Maller
A.H., Zentner A.R. (2008) Merger Histories of  Galaxy
Halos and Implications for Disk Survival. The Astrophysical Journal
683, 597-610. \url{http://arxiv.org/abs/0711.5027}
 \item[]
van den Bergh, S. (1998) Galaxy Morphology and Classification. Cambridge
University Press
 \item[]
Wiens J.J. (2001) Character Analysis in Morphological Phylogenetics:
Problems and Solutions. Syst. Biol. 50, 689--699.
 \end{description}

\end{document}